\documentclass[9pt,twocolumn,twoside]{pnas-new}
\templatetype{pnasinvited} % Choose template 
\title{Agent Based Computational Model Aided Approach to Improvise the Inequality-Adjusted Human Development Index (IHDI) for Greater Parity in Real Scenario Assessments}

\author[a,1]{Pradipta Banerjee}
\author[a,2]{Subhrabrata Choudhury}

\affil[a]{National Institute of Technology Durgapur, India}

\leadauthor{P. Banerjee and S. Choudhury} 

\authorcontributions{}
\authordeclaration{}
\equalauthors{}

\correspondingauthor{\textsuperscript{1}To whom correspondence should be addressed. E-mail: pradiptapb\@email.com}


\begin{abstract}
To design, evaluate and tune policies for all-inclusive human development, the primary requisite is to assess the true state of affairs of the society. Statistical indices like GDP, Gini Coefficients have been developed to accomplish the evaluation of the socio-economic systems. They have remained prevalent in the conventional economic theories but little do they have in the offing regarding true well-being and development of humans. Human Development Index (HDI) and thereafter Inequality-adjusted Human Development Index (IHDI) has been the path changing composite-index having the focus on human development. However, even though its fundamental philosophy has an all-inclusive human development focus, the composite-indices appear to be unable to grasp the actual assessment in several scenarios. This happens due to the dynamic non-linearity of social-systems where superposition principle cannot be applied between all of its inputs and outputs of the system as the system's own attributes get altered upon each input. We would discuss the apparent shortcomings and probable refinement of the existing index using an agent based computational system model approach.
\end{abstract}

\dates{}
\doi{\url{}}

\begin{document}

\maketitle
\thispagestyle{firststyle}
\ifthenelse{\boolean{shortarticle}}{\ifthenelse{\boolean{singlecolumn}}{\abscontentformatted}{\abscontent}}{}

\noindent The quest for social system models to establish a harmonious state of affairs and stable egalitarian humane living conditions among all the people of the society has been the most coveted question in societies since the emergence of civilisation. However, it has been observed throughout the passage of time that seemingly unbridgeable gaps exist between the expectations generated from several rational beliefs about the social systems and the actual evolved outcomes in the corresponding systems. These expectation-outcome gaps further expand the disruptions in the social system structures. Thereby the resultant social system scenarios become even more incompatible for being assessed using those existing conventional rational beliefs. This leads to greater divergence of opinions, which merely results in generation of new sets of theorised beliefs that have similar limited capabilities. In more general terms, a mismatch between the driving forces of rapid changes in socio-economic activities of the world and the decision-making structures of most societal institutions gets evident in due course of time. These phenomena happen as these institutions across time, geographic and cultural domains appear to be ill-equipped for managing the subversive emergent processes in the social system. These shortcomings fail to subdue the subversive phenomena, which continue to exert direct/indirect influences on the lives of individuals and their communities in the society resulting in the societal policy expectation-outcome gaps. Thereby assessment of true amount of development becomes even more challenging. In this paper, we try to highlight the problems in effectively adjudging the amount of human-development through the existing Inequality-Adjusted Human Development Index (IHDI). Subsequently we propose Agent Based Computational methodologies for probable improvisations of the composite-index to overcome its shortcomings. 

\section*{Related Work}
Human Development Index (HDI) developed by Amartya Sen and Mehbul ul Haq in 1990 \cite{UNDP} has been the most significant composite index till now to be credited for credible assessment of human well-being. Amartya Sen has remained concerned about the crudeness of the HDI from the initial phases \cite{Amartya1}. However he was later on convinced by Mahbub ul Haq's argument justifying the enormous significance of the index that could bring about a fundamental change in the perspective view of Human Development, competing directly with the artificial statistical measuring metrics like the crude GDP per capita numbers and  Gini coefficients that remained predominant in development thinking. The HDI is based upon the geometric mean of Life Expectancy, Education and Income. The composite index was further adjusted for inequality in each of its dimensions to develop the IHDI in 2010 \cite{IHDI1,IHDI2}. Though it did refine the composite index but the inherent shortcomings in the index rather practically defies the core philosophy behind the Capability Approach as perceived by Amartya Sen \cite{sen1999development} and later on by Martha Nussbaum \cite{nussbaum2011}. Amartya Sen's Critiques \cite{sen1979} of Act-Consequentialism, Welfarism, Sum Ranking in Utilitarianism and Resourcism highlight the humanitarian outlook missing in welfare economics. Thereafter Martha Nusbaum developed a very comprehensive systematic, and influential capability theory of justice. She aims to provide a partial theory of justice (one that does not exhaust the requirements of justice) based on dignity, a list of fundamental capabilities, and a threshold \cite{nussbaum2011}. Other than just individual physical capabilities, life expectancy, health, she even takes into account, rights for protection of an individual's self-respect, to participate in societal processes, to think and feel freely. In her capability theory, she also considers rights to express emotions freely, to be attached with other individuals, to be able to play and have fun. Even capabilities to be able and sensible to live with concern for and in relation to animals, plants and the world of nature along with having freedom to control over one's own environment, have been taken into account.
\par
However, Amartya Sen has not thereof specified any list of all the capabilities which are important and how they need to be distributed, as to him these are political decisions for the society itself to decide. Both capability theorists and external critics expressed related concerns about the institutional structure of the Capability Approach. The Rawlsian social justice theorist, Thomas Pogge in \cite{pogge2002can} also have raised a few pertinent questions. He accentuated as how should capabilities be weighted against each other and non-capability concerns? For example, should some basic capabilities be prioritized as more urgent? What does the Capability Approach imply for interpersonal equality? How should capability enhancement be paid for? How much responsibility should individuals take for the results of their own choices? What should be done about non-remediable deprivations, such as blindness? Amartya Sen's main response \cite{sen2010} to such criticisms has been to admit that the Capability Approach is not a theory of justice but rather an approach to the evaluation of effective freedom. Amartya Sen also acknowledged the fact that it's easier to gather information of some capabilities than others \cite{}. Thereby it is evident that HDI and IHDI too, does not fully reflect the scope or methodology of the Capability Approach. IHDI does not intrinsically account for the relative geographic, socio-political, demographic, cultural differences between regions having same indices.  Furthermore, due the psycho-physiological heterogeneity of human-beings, the HDI/IHDI assessment appear to be unable to bring-out the real picture of human development defying the designing philosophy of the index.

\section*{Motivation and Problem Statement}
The drawbacks of this composite-index can be elaborated through the following case scenarios. Saudi Arabia belongs to the very high human development list having HDI rank 39, as per Human Development Report 2018 \cite{IHDI}. Still a government sponsored Mobile App ``Absher" is prevalent in the country \cite{Absher} revealing the grave scenario of gender inequality. Women of Saudi Arabia and Middle East countries, many of which belong to the very high human development group, are also found to have Vitamin D deficiency \cite{veil1,veil2,veil3,veil4} in general, where one of the key reasons identified for this phenomenon is their predominant veil regime. Four out of the top five nations having the maximum number of suicides per year \cite{sui}, Lithuania (28) \footnote{The value inside the braces indicates the IHDI ranking as per  Human Development Report 2018.}, Russia (34), South Korea (23) and Belarus (30), belong to the very high human development group as per IHDI 2018 \cite{IHDI}. Lower IHDI ranked countries like India (95), Iraq (83), Azerbaijan (47) do have much better Inequality-adjusted Income Index ranks, in spite of having much lower Inequality-adjusted education and life expectancy ranks, than the corresponding much higher IHDI ranked countries Panama (61),  China (56) and Russian Federation (34)  respectively as per Human Development Report 2018 \cite{IHDI}. Simply errors due to the quality of obtained statistical information by the United Nations cannot serve for a justified explanation. The three dimensions (education, income and life-expectancy) of HDI/IHDI seem to be independent of each other and also are unable to project the actual scenario of Human Development both individually or collectively. For example, how can Human Development be high when one half of the society is under the covers of suppression or unknowingly suffering from high Vitamin-D deficiencies, leading to several critical health problems \cite{Vit1,Vit2,Vit3,Vit4,Vit5}, simply due to some rigid social customs, or how is this possible that enriched human capital due to education is having much lower incomes than the uneducated ones.

\par
\begin{figure}[htp]
   \centering
   \subfloat[Pattern of accumulation of private wealth from Pre-World War era in Germany, France and Britain]{\label{rev_sol}
      \includegraphics[width=.50\textwidth,height=4cm]{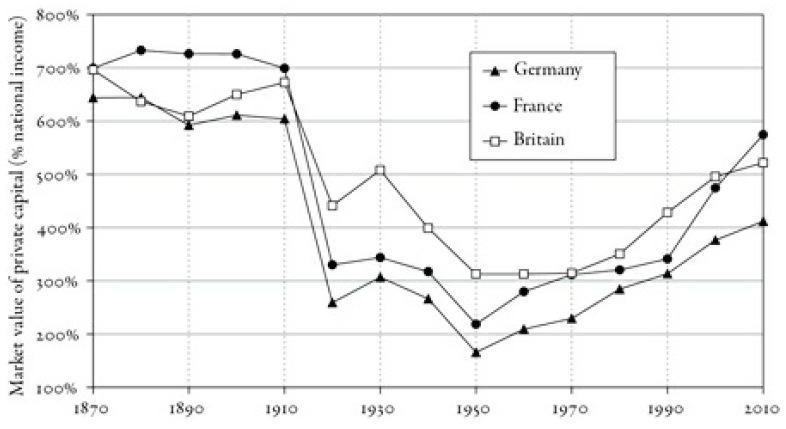}}
 
     \subfloat[Pattern of accumulation of private wealth from Pre-World War era in US]{\label{rev_sol}
      \includegraphics[width=.50\textwidth,height=4cm]{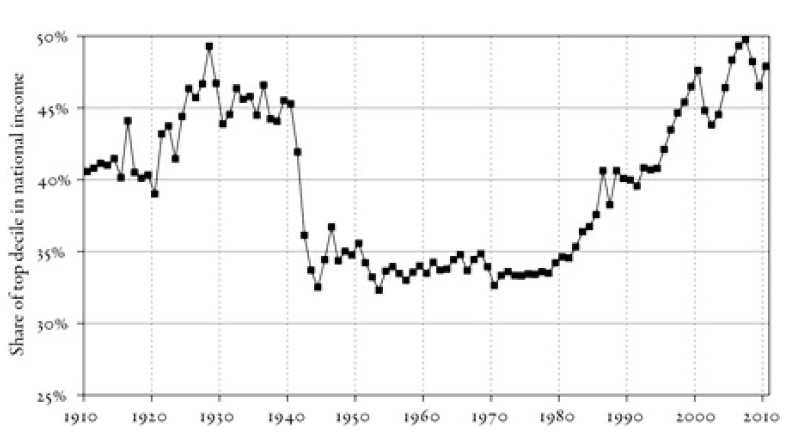}}
   
   \subfloat[Comparison of patterns of accumulation of private wealth from Post-World War era]{\label{rev}
      \includegraphics[width=.50\textwidth ,height=5cm]{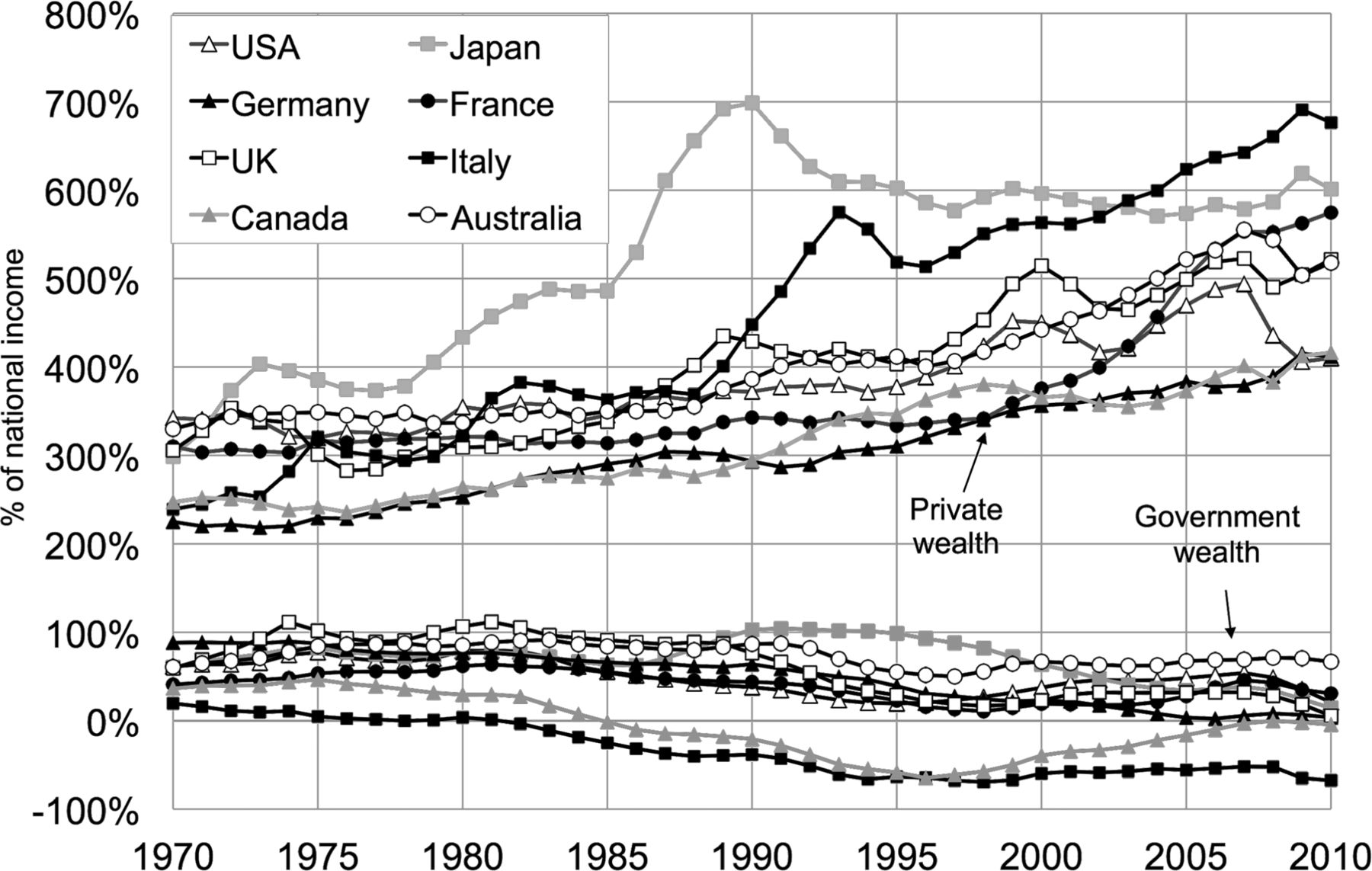}}

  \subfloat[Comparison of patterns of accumulation of private wealth with China from Post-World War era]{\label{rev_sol}
      \includegraphics[width=.50\textwidth,height=5cm]{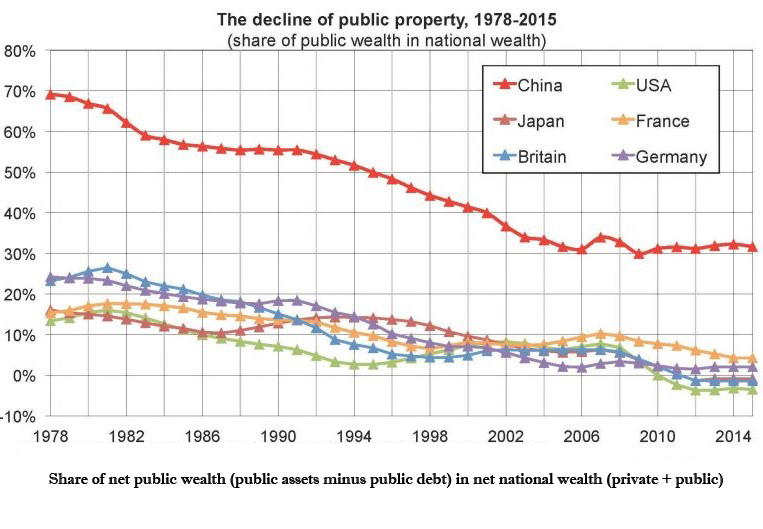}}

   \caption{Illustration of the pattern of accumulation of private wealth from Pre and Post World War era \cite{Piketty}}\label{bs1}
\end{figure}

 
Delving in this context, pertinent questions about the socio-economic system arise regarding the Accumulation of Private Capital and widespread rising inequality in distribution of wealth. In this regard the comparison charts in Fig. \ref{bs1} Thomas Piketty's ``Capital in the 21st Century" \cite{Piketty} raises such salient questions.

World Wars have largely contributed positively in decreasing the concentration of private capital in the developed countries. This can be understood as any type of catastrophe natural or wars largely destroy properties of all property owners, big or small. However, the prominent question that arises is why does the process of accumulation of private capital rise again with consequent development as evident from the charts in Fig. \ref{bs1}. It is also evident that post Industrial Revolution, returns to human capital garnered more returns as compared to income from private wealth such as land and inherited property. Apparently many avenues of earnings by the virtue of personal skills overtook the income from private capital. In this process several age-old entrepreneurs perished and several new entrepreneurs came up to the stage from nowhere. The transformation processes of riches from rags and vice-versa became eminent. This lets us have an insight in the process of accumulation of private capital. That is the capital has been changing hands during various stages of economic growth but the accumulation of capital intensifies barring certain transient intersections of change of hands. Furthermore, another question comes to the forefront regarding the concern for widespread increasing inequality in distribution of wealth. If the topmost decile possesses most of the national wealth, then for the rest of the population intra and inter decile variance in wealth tends to be much lower. Then why there are grave concerns of unstability for societies suffering from such inequalities even when the subsistence wage is also many times lower in the modern world than it was in the preindustrial period \cite{gregory} due the advances in science and health technologies. Even-though the intra and inter decile, below the topmost decile, possess similar patterns of variations in accumulation of private capital are observed, the variances therein are low. Yet the expectation - outcome gaps for the individuals and collective groups of individuals seem to result in an unstable chaotic scenario. 
\par
These occurring phenomena cannot be explained by comprehensive consideration of objective and subjective mental metrics, as even only subjective psychological influence metrics can outrun such comprehensive analysis. Furthermore, the accumulation of private capital patterns highlight the negative effects of `adaptive preferences' \cite{sen1985commodities}, which has been not considered while designing assessment indices, wherein the poor on becoming richer gradually start to exhibit the capitalist behaviour of the richer resulting in more accumulation of private capital and deprivation of the weak. This happens due to the continuous strive for growing individualistic aspirations of individuals or groups of individuals in the society. In reality individuals and groups of individuals rely, not only on one's own capacity and capability, but also tries to gain one-sidedly by using others' capabilities which relates to the primitive capitalistic instinct.

\par
STEM (Science, Technology, Engineering and Mathematics) based formal education has little in their offing towards the essence of ethics, civic sense and moral science having a narrow focus of learning in the long run. Knowledge of science and technology can have infinite boons upon the human race but practically lacks the breeding spree of inherently moralised humans in spite of having the basic intent for the same. To add to it merely more education also does not relate to more human capital, as objectively perceivable individuals in same condition would get enriched to different levels due to variation in their individual cognitive and behavioural capabilities. Individuals have their bounded set of capabilities, where physically and mentally complete unproductive relates to the lower bound and having productivity to its topmost level serves as the upper-bound. Their capabilities and efficiencies also vary within this range depending on the endogenous parameters effected by age, disease, interpersonal relationships, stress levels due to exogenous socio-political and natural environment leading to imbalance in levels of cortisol and seratonin in one's body. Remarkable variations in cognitive and physiological efficiencies have been observed to be impacted by the cortisol \cite{cortisol1,cortisol2,cortisol3,cortisol4,cortisol5} and seratonin imbalances \cite{seratonin1,seratonin2,seratonin3}.  Moreover the ability to cope up with the stress or the proneness to get stressed too are depended on several genetic factors \cite{serge1,serge2,serge3} though the specific genes contributing to these phenomena are still being discovered. The assumption of all individuals educated to the same level would possess the same amount of knowledge suffers from the same shortcoming of `sum-ranking' \cite{sen1999development} and such assumptions would merely result in a high level of literacy rather than the expected educational value. This is because of their variations in cognitive and behavioural abilities. Imposition of \textit{fait accompli} regarding reliability precepts upon blood tied or socially-contracted relations are generally predominant over equivocal faith upon the support system of the human community around. As a result, even though it is realisable that an individual's life from its initiation bears no predefined fixed set of routes to be traversed, but still fixed interpersonal expectations and aspirations based on such precepts get prevalent. This too impacts the general capabilities of individuals through inter-personal relations. Learning to inculcate the basic morals of communityhood remains a challenge for attaining desirable socio-economic and ecological optimal equilibria.

\par
Sheer focus on expected life expectancy can be misleading, in the process of giving stress upon artificial support systems and medication for long life but void of expected quality of life. The need should have been to emphasize on a quality of life that would be satisfiable to the all-inclusive society having at least the average life expectancy that encompasses a handsome number of years post the generic universally accepted retirement age. Desirable stable egalitarian equilibrium can only be attained when there is no inflation and there is no deflation too, indicating that commodity or service prices have reached the values beyond which it will be unsalable and below which it would generate inadequate revenue for paying off other input prices. This indicates the optimal strategy selection in classical game theoretic approach for attaining a stable Nash Equilibrium. This can happen only when prices, incomes and the demographic distribution (within specific bounded size limits) as per age-capability are stable and tend to be constant, while considering the available amount of natural resources to be limited. The reason behind this is whatever human or material resources are developed, all correlates to utilisation of naturally available resources in some form or the other, and these utilisation does not generally relate to restorable exhaustion of resources. This price-income setting can be achieved only when the number of people retiring from income is equal to the number of people starting to earn with the prior condition that the total population is constant. Such scenario demands that the gross loss in the desirable capability of the society due to death-rate should get compensated by the gain in capability due to birth-rate. Further now this desirable setting demands some basic rethinking in the individual choices and underlying societal norms regarding reproduction and familyhood while reforming the mindset towards communityhood. The Population/Age-Sex pyramid needs to be stationary tubular rather than expansive pyramidal or constrictive pyramidal in shape with feasibly narrow width so as to keep natural and physical resources in affluence rather than systematically depleting it. This is crucial because the feelings of insecurities associated with the scarcity principle induces capitalistic behaviour and thereafter the capitalist mentality in return propels the feeling of scarcity principle \cite{scarce1,scarce2,scarce3} creating a systemic vicious feedback loop.

\par
These facts also justify that any constituent principal-subsystem of the composite socio-economic system can adversely affect and even negate the positive utilities of another principal-subsystem resulting in overall failure of the socio-economic system. Only a stable setting for securing individual expectations and aspirations, where all the individuals in the social system have low variance in the upper and lower bounds of individualistic expectations and aspirations, can lead to stable and egalitarian setting for the society. The individual capabilities does not only rely on their physical abilities but also on their behavioural and cognitive aspects. These behavioural and cognitive abilities are completely heterogeneous having several dependencies on their individual genetic structures \cite{gest1,gest2,gest3,gest4,gest5} with some directly inherited and some being randomly evolved. As observed they also correlate to mental and cognitive disorders, posing more sets of capability deficiencies that cannot be compensated by merely providing direct physical resource incentives which on the other-hand may just turn out to be futile exercises resulting in disproportionate wastage of resources. As per general observations individual brain structures are infinitesimally different, due the varied genetic structures but even the slightest change can result in exhibition of grossly different characteristics \cite{}. Therefore each human being (the individual social-agent) relates to a scenario having different intrinsic neural networks where same inputs to each of the networks would result in varied outputs through each of the different networks correspondingly. So we have to modulate the inputs such that the self adaptive neural networks slowly evolve in such manner that same inputs confer to similar outputs. Oded Galor and Quamrul H. Ashraf in \cite{macrog1} and their joint research \cite{macrog2} with Cemal Eren Arbatli, Marc Klemp too highlighted that the varying degrees of genetic diversity in the population of a place is responsible to a great extent for the quality and amount of that particular place's socio-economic development. The cultural paradigms of different places too evolve from such diversity.

\par
 Amartya Sen raised his concerns regarding `act-consequentialism' \cite{sen1985commodities}, where he emphasized that not only the outcome is important but the process through which that outcome is achieved needs to be accounted for in the comprehensive analysis of social systems. Thus it is quite apparent from the above observations that in addition to this concern, the non-uniformly varying quality and feedback aspects of such processes, which are changing the initially expected behaviour and outcome patterns, restructuring the social system scenario, are of prime importance. Quantification of the qualitative aspects of social actors and a framework to correlate the quantitative analysis with empirical economic analysis remains a challenge. To the best of our knowledge even if the theoretical aspects were analysed \cite{}, the analyses were not well equipped to predict the future real outcomes with precision for any given policy. Moreover, these analytical studies in general have been limited to single level of exogenous influences in the social system. They do not account for the multilevel convoluted structure of influences across the time domain amongst contextually related social agents/agent-groups, which results in the expectation-outcome gaps of socio-economic policies. This happens because, in reality, a constituent (individual or collective institution of social agents) of the social structure, not only does receive/gather some input from related social-subsystems for delivering some output, but it also undergoes self-structural changes wherein the same input to it in future would produce some different output compared to the one yielded now. Thus these factors poses serious challenges in designing realistically analogous composite assessment indices for human development.
 
 \section*{Proposed Approach}
\begin{figure}[htp] 
\centering
\centerline{\includegraphics[width=3.6in,height=7cm]{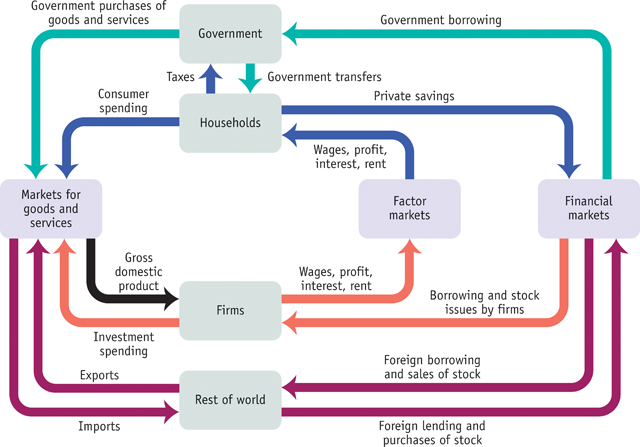}}
\caption{An Example of Conventional Macro Economic Model\cite{conve}}
\label{Conventional Model}
\end{figure}

 In this paper we would propose an approach to improvise the introspection of the macro aspects arising from the micro behaviour with greater realistic analogy. For this purpose we view all individuals, collective groups of individuals, institutions as social-agents. We consider all social-agents having heterogenous capability, expectation attributes. The interplay between these social-agents result in a dynamic composite macro behaviour. This macro behaviour has different dimensions having varying effects upon different activities in the society. Traditionally the socio-economic systems were studied in isolation as shown in one of the general conventional diagrammatic representation of socio-economic systems in Fig. \ref{Conventional Model}.

\begin{figure}[htp] 
\centering
\centerline{\includegraphics[width=3.6in,height=7cm]{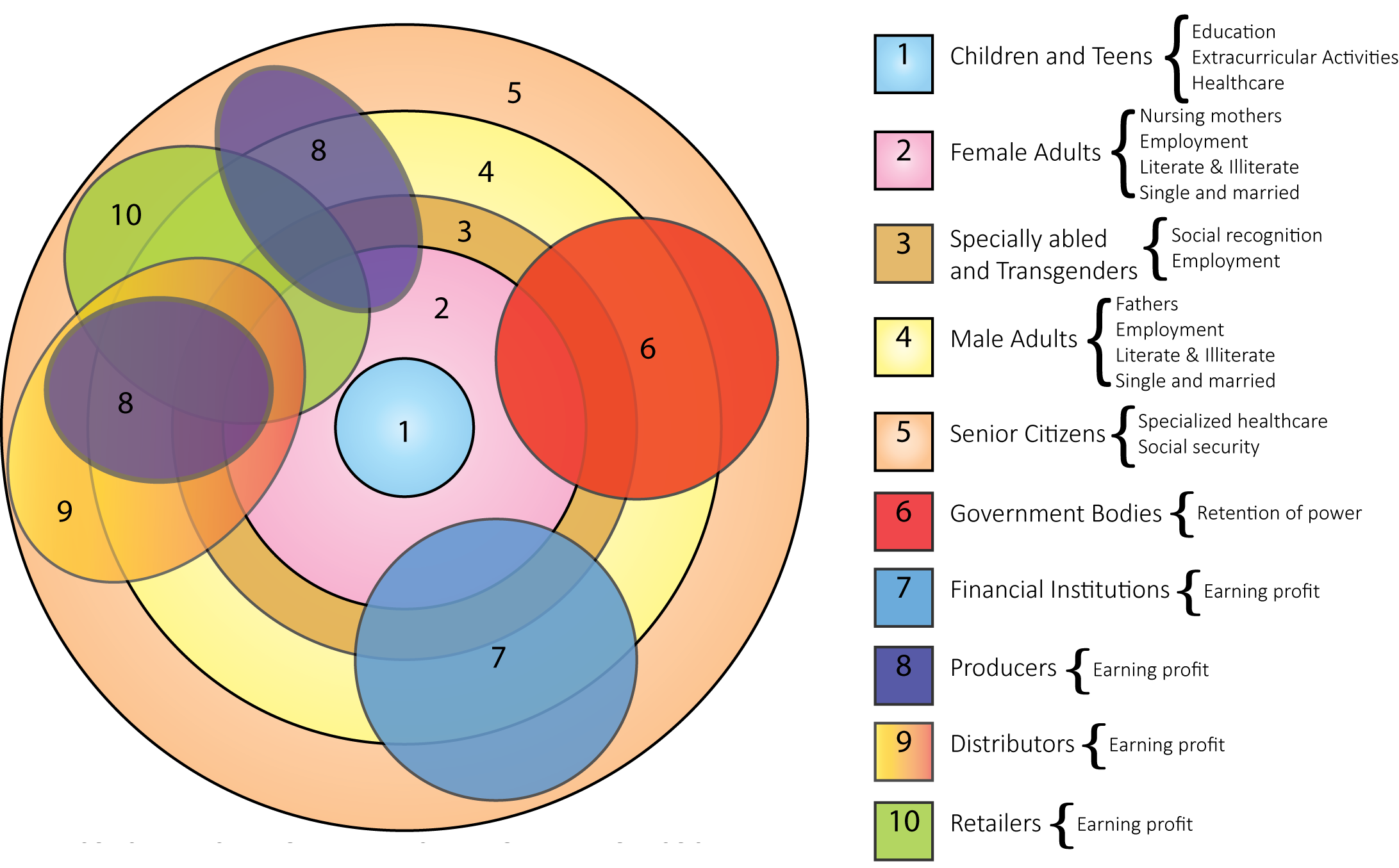}}
\caption{The Consumer Set Model Representing the Entire Society}
\label{Consumer Model}
\end{figure}

 We view the entire society as the composite socio-economic system comprising of the underlying principal-subsystems of basic activity processes in the society. We have conceptualised the socio-economic system modelling approach by concurrently incorporating the principal inter-related subsystems deriving realistic analogy of the actual society. All social-agents including individuals (human-beings) or groups in collective form, do (need to, expect to and wish to) consume some of the natural and social resources irrespective of the fact whether they produce or contribute something to the society and thereby they do have a considerable effect upon the society and it's resources. Hereby we consider all the constituent agents of the society as consumers. The diagrammatic representation of the complete society as The Consumer Set Model is illustrated in Fig. \ref{Consumer Model} elaborates the sectional classification of the agents based on their social existence. At present we would restrict our discussion to the macro behaviour involving the interplay of these principal-subsystems, which is actually the aggregate influence outcome of the micro-constituents (social-agents) of the subsystems. These principal-subsystems too influence each other thereby making the complete dynamic nature of the socio-economic activities inherently evident. 

\begin{figure}[htp] 
\centering
\centerline{\includegraphics[width=6cm,height=6cm]{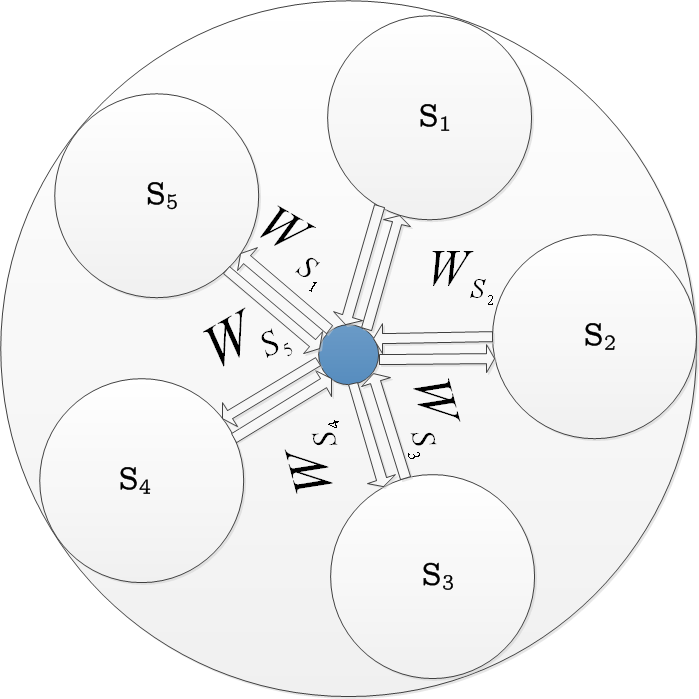}}
\caption{The Set \textbf{S} representing the Composite Socio-economic System contains the Interplay of Influences between the Principal-Subsystems in the Society at time $t$ ($W_{S_{i}}$ relates to the performance metrics of the principal-subsystem $S_{i}$ at time $t$)}
\label{System}
\end{figure}

We have analysed that there are $5$ principal-subsystems (based on analytical studies \cite{stud1,stud2,stud3,stud4,stud5}) of the society which actually determines all the parameters of the socio-economic activities. These are - 1) Comprehensive Education System 2) Health-care and Nutrition Access 3) Public Insurance  and Micro-financing Frameworks 4) Human Security and Legal Systems 5) Income Avenues, Technological Growth and Demographic Transition Management. These principal-subsystems are interrelated and influencing each other simultaneously evolving the society which is represented in Fig. \ref{System}.

\par Let us consider $\textbf{S}$ to be the set comprising of all the 5 principal-subsystems in the composite socio-economic system such that $\textbf{S}=\{ S_{1},S_{2},S_{3},S_{4},S_{5} \}$. Here $S_{1},S_{2},S_{3},S_{4},S_{5}$ represent the principal-subsystems - 1) Comprehensive Education System 2) Health-care and Nutrition Access Platform 3) Income Avenues, Public Insurance and Micro-financing Framework, 4) Human Security and Legal Systems 5) Technological and Demographic Growth/Transition Management respectively. Each principal-subsystem is interrelated with each other such that the fixed utility weight factor of $S_{i}$ resulted by $S_{j}$ is $U_{ij}$ and the dynamic relationship strength of $S_{j}$ incident on $S_{i}$ is $R_{ij(t)}$ at a time instance $t$. Here $t$ represents the timestamp in the concerned timeseries for analysis. Each $S_{i}$ has a total utility weight parameter corresponding to each specific time $t$ in the composite socio-economic system as $W_{{S}_{i}(t)}$. This total utility weight parameter, $W_{{S}_{i}(t)}$, relates to the performance metrics of the principal-subsystem $S_{i}$ at time $(t)$, for example Learning Rate may be considered as the performance evaluation metric for the Comprehensive Education System. It can be represented as -\\

\begin{equation} \label{eq:1}
W_{{S}_{i}(t)} = \sum_{j=1}^{\mid \textbf{S} \mid} R_{ij(t)} \times U_{ij} \mbox{,    where  } {\mid \textbf{S} \mid} \mbox{ is the cardinality of the set \textbf{S} } \\
\end{equation} 
\\
$R_{ij(t+1)}$, dynamic relationship strength, is the measure of the ratio of change total utility weight parameter of the principal-subsystem $S_{i}$ with respect to the change in that of $S_{j}$ in the previous cycle. The values of dynamic relationship strength, $R_{{ij}(t+1)}$, of $S_{j}$ incident on $S_{i}$ at time $(t+1)$ is calculated as  - \\

\begin{equation} \label{eq:2}
 R_{{ij}(t+1)}= \begin{cases} \mbox{    } 0, \mbox { if }  (W_{{S}_{i}(t)} - W_{{S}_{i}(t-1)}) \veebar (W_{{S}_{j}(t)} - W_{{S}_{j}(t-1)})=0 \\ \\\mbox{    } R_{{ij}(t)}, \mbox { if }  (W_{{S}_{i}(t)} - W_{{S}_{i}(t-1)}) = (W_{{S}_{j}(t)} - W_{{S}_{j}(t-1)})  \\ \\\mbox{    } {{\mid\frac{(W_{{S}_{i}(t)} - W_{{S}_{i}(t-1)})}{(W_{{S}_{j}(t)} - W_{{S}_{j}(t-1)}) \times R_{{ij}(t)}}\mid }}^{sgn(\frac{(W_{{S}_{i}(t)} - W_{{S}_{i}(t-1)})}{ (W_{{S}_{j}(t)} - W_{{S}_{j}(t-1)}) \times R_{{ij}(t)}})}, \\\mbox { otherwise }\end{cases} \end{equation} \\ 

 We need to derive $W_{{S}_{i}(t)}$ and $W_{{S}_{i}(t-1)}$ from statistical data for a given time. Initial $R_{ij(t)}$ values for all the principal-subsystems need to be adjusted/tuned and set in the normalized scale of $0$ (no relation) to $1$ (directly proportional influence) based on analytical studies. This needs trial and error tuning to check which initial settings, upon applying the known policy function $P_(t)$ during that specific time instance, that yields closer to known real values for the subsequent time instance during the simulation of the system model. This approach is in consonance with the concerns for the precarious results that may arise, while working with social systems, due to the gaps between moderately accurate simulation results and the actual events, as detailed by Wallach in \cite{hanna}. From Equ. \ref{eq:1} we have a set of equations in this scenario where the values of $W_{{S}_{i}(t)}$ and $R_{ij(t)}$ are known and the corresponding fixed utility weight factors $U_{ij}$ are the unknowns as shown in Equ. \ref{eq:3}.

\begin{equation} \label{eq:3}
W_{{S}_{i}(t)} =  \sum_{j=1}^{\mid \textbf{S} \mid}\sum_{k=1}^{\mid \textbf{S} \mid} R_{{jk}(t)} \times U_{jk}; ~~ \forall ~i=1~ to ~ \mid\textbf{S}\mid \end{equation} \\

We can solve the set of these equations, each of which here represents an underdetermined system, using the method of Lagrange multipliers to derive the values of $U_{ij}$. Conjointly using these initial values with the performance metrics derived for each principal-subsystem in future cycles, we can evaluate as to the level of sensitivity each principal-subsystem is having towards its corresponding influencer principal-subsystems in a certain scenario. For this purpose we construct a matrix having the principal-subsystems and their corresponding principal-systems as influencers. A toy example with random values has been shown in Table \ref{System Weight}. From this example table we can infer that the normalized value of the relationship strength of influence of Comprehensive Education System ($S_{1}$) incident upon the Human Security and Legal Systems ($S_{4}$) is $0.3$ whereas Human Security and Legal Systems ($S_{4}$) has no influence relation upon Comprehensive Education System ($S_{1}$) as the corresponding value is $0$ at time $t$.

\begin{table}[]
\centering
\begin{tabular}{c|c|c|c|c|c|}
\cline{2-6}
                         & $S_{1}$  & $S_{2}$  & $S_{3}$  & $S_{4}$  & $S_{5}$  \\ \hline
\multicolumn{1}{|c|}{$S_{1}$} & 1 & 0.9 & 0.1 & 0.3 & 0.2 \\ \hline
\multicolumn{1}{|c|}{$S_{2}$} & 0.3 & 1 & 0   & 0.2 & 0.4 \\ \hline
\multicolumn{1}{|c|}{$S_{3}$} & 0.4 & 0.6 & 1 & 0   & 0.1 \\ \hline
\multicolumn{1}{|c|}{$S_{4}$} & 0   & 0.5 & 0.2 & 1 & 0   \\ \hline
\multicolumn{1}{|c|}{$S_{5}$} & 0.7 & 0.6 & 0.2 & 0 & 1 \\ \hline
\end{tabular}\\ 
\caption{Influence Matrix of Dynamic Relationship Strengths $R_{{ij}(t)}$ between the principal-subsystems at time $t$}
\label{System Weight}
\end{table}

\par 
The basic algorithm for the generation of these matrices at any time instance $t$ is shown in Algorithm \ref{Algo}. This would further enable us to design policies with specific quantitative emphasis on principal-subsystems for achieving the desired state of the composite socio-economic system at time $(t+1)$, 

\begin{algorithm}[H]
\SetAlgoLined
 \For{$i = 1~to~ \mid\textbf{S}\mid$ }
 {
 initialize: $W_{{S}_{i}(t-1)}$, $W_{{S}_{i}(t)}$;   \tcp*[h]{\footnotesize{Fetch from performance analysis metric statistics of $S_{i}$ at time $(t-1)$ and $(t)$ respectively}}\\
 \For{$j = 1 ~to~ \mid\textbf{S}\mid$ }
 {
  
  Assume $R_{{ij}(t-1)}$ value between $0 ~to~ 1$; \tcp*[h]{\footnotesize{Based on statistical analysis}}\\
  Approximate $U_{ij}$;\\ 
 \While{$P_{t-1}(R_{{ij}(t-1)},U_{ij}) \neq W_{{S}_{i}(t)}$} 
  {\tcp*[h]{\footnotesize{$P_{t-1}$ is the known policy function adopted at time $(t-1)$}}\\
  Tune $R_{{ij}(t-1)}$\; 
   }
   Calculate $R_{{ij}(t)}$ using $R_{{ij}(t-1)}$;\\
   
   }
  }
  Form the Influence Matrix;
\caption{Formation of Influence Matrix}   
\label{Algo}
\end{algorithm}

\par 
Cumulative super-positioning effects of inter and intra principal-subsystems in the prior steps are embedded in the performance metric of each principal-subsystem in the present step. From this matrix we can further use feature extraction tools, like Principal Component Analysis, to find out the most influential principal-subsystems in the composite socio-economic system at a given time $(t)$ to design requisite policies for the particular scenario at that time. We can further analyse how the inter-relational strengths and total utility weight parameters of principal-subsystems self-adjust under the influence of it's own performance and impact of other principal-subsystems. From these observations we can further infer while designing policies, given certain initial scenarios, what would be the right proportionate emphasis on specific principal-subsystems to fetch the desired outcomes by overcoming the undesired impacts of the underlying influences in the composite socio-economic system. 

\subsection{Incorporating the proposed approach with IHDI and its analysis}

 IHDI serves to be the foundational human development composite-index available and to the best of our knowledge, is the most reliable and widely accepted index projecting human development growth of regions. We improvise it using it as the comparative reference to formally describe the level of actual Human Development and also project the variation between the qualitative amount of principal-subsystems' development and same ranked IHDI regions. For this purpose, the macro indicator can be related with the aggregate analysis of the performance metrics of the considered principal-subsystems. We introduce the concept of a quality proportioning coefficient $Q_{c(t)}$ in Equ. \ref{eq:4}. $Q_{c(t)}$ accounts for the geographic, socio-political, demographic, cultural differences which are the underlying factors affecting the performance parameters of the constituent principal-subsystems of the society. \\

\begin{equation} \label{eq:4}
Q_{c(t)}\times IHDI_{(t)}=  \sum_{i=1}^{\mid \textbf{S} \mid} \frac{W_{{S}_{i}(t)}}{{\mid \textbf{S} \mid}}\end{equation}\\

 Herein we would also highlight the fact that a particular policy for a region is not proportionately appropriate for another region even if they share the same IHDI unless their $Q_{c(t)}$ values tend to be same.  Human species like all other species too have variance in their breeds accounting from their places of origin based on their natural surroundings, habitual and eventually cultural differences. These differences also sum up to enormous impacts onto the policy designs. Consideration of such impacts, which has been lacking in the quantitative analyses of conventional socio-economic theories, has been accounted for through this approach. Every crisis scenario has to be treated individually taking into account the current status and the known data of preceding scenarios for that specific region. It is also evident that a policy pattern cannot be static for a specific region under the impact of networks of underlying influences between the principal-subsystems of the society and needs to be restructured periodically based on the evolved scenario observations over feasibly short intervals of time to overcome the evolving undesirable impacts of the intangible influence networks. IHDI is considered here because it serves as the comparative composite index which expresses the levels of income, education and life expectancy. And we try to relate these levels with the actual quality of life therein. In the ideal case both should be homologous but as we have observed, that due to the sheer heterogeneity amongst the social-agents (individuals/groups) having time varying nature and interactive adaptive influences upon one another, the levels of IHDI values does not represent the actual quality of life. Through this approach we try to relate these two different attributes, that is, the observed levels to the actual quality yielded considering the underlying complex network of influences. We introduce the concept of the quality proportioning coefficient, $Q_{c(t)}$, to account for the propotionality of the index of a region to the quality of life there. $Q_{c(t)}$ reflects the actual quality of the functional status of the principal-subsystems and the micro constituents, the social agents, therein relaying the true picture of the composite socio-economic system there. Ideal objective values for $Q_{c(t)}$ and IHDI should be $1$, but any value, for both $Q_{c(t)}$ and IHDI, above $0.9$ having a stationary or increasing trend can be considered as indicator of a realistically satisfiable quality of life. This can be achieved via evolutionarily cultivating epigenetic transformations (inducing behavioural changes through external enviornment stimuli, through adaptive policies, on an epigenetic level) \cite{epig} for achieving expected functionality bounded by desired upper and lower behavioural thresholds. This would result, with much high probability, in the expected external behavioural manifestation after rounds of biological evolution through the intrinsic attributes (natural genetic behaviour) of social-agents. Another important aspect apparently emerges out whilst considering this probable improvisation to IHDI is that if we deal with IHDI attributes of smaller regions such as districts, states in place of countries as a whole, we would achieve better accuracy in observing and designing required policy frameworks. This is because consideration of larger regions often under-mines the varying inter-regional cultural, demographic, geographic, socio-political fall-outs. The proposed system model approach provides the platform for establishing greater correlation between the macro and micro views of the social system. Thereby this has the potential to enable us for designing and tuning truely adaptive synchronised polices for any existing scenario at a specific time instance. This approach for improvising the existing IHDI composite human development index may fetch desired outcomes overcoming the non-desirable arbitrary impacts of the underlying influence networks in the society.
 
\section*{Conclusion and Future Works}
Our approach towards reducing the expectation - outcome gap is by analysing the interplay between social-agents using Agent Based Computational approach. Through this proposed approach we tried to bring the insight about the convoluted influence of influences between the principal-subsystems of the society which takes shape as a complex time folded network of influences. The structural approach outline for the functioning of the micro constituents of the composite socio-economic system, the social agents, for designing the individual principal-subsystem models can be conceived from the similar pattern of the proposed approach. The primary requisite for this is the true assessment of the parameters of a crisis scenario. The analysis should account for the convoluted multiple degrees of influences (influence of influences) of adaptive social-agents in the composite socio-economic systems, to correlate the observed deviations of macro level data from estimated values. We have tried to analyse and modularize the social system into perceivable self restructurable principal-subsystems, which can be aptly interlinked in a comprehensive manner to represent the realistic scenario. Such models should also encompass the time dependent multi-layered influence of the social-agents. This would enable us to aptly analyse the dynamic adaptive complex networks with multi-level influences of social-agents, which actually outweigh their embedded impacts in long runs. Such an approach is in contrast to the classical game theoretic approach that accounts for single level of influences with binary choices of co-operate and deflect. Designing models based on this approach would enable us to maintain analogy with the real socio-economic system having un-constrained dynamic sized adaptive strategy sets accounting for the varying gradients of co-operate/defect, heterogeneity and size of agent population. Constructing such models for the individual principal-subsystems remains the future scope of work in this context. The task once accomplished would enable us to represent the all-inclusive society provisioning the platform for both top-down and bottom-up views for observing the evolutionary socio-economic system. This would account for the macro-micro, reverse macro-micro linkages and the embedded networks of influences formed therein. Such frameworks would let us prevent deprivation of any deserved from the requisite human and physical resources and at the same time put a check on the misutilization and under-utilisation of  such resources. Thereby such frameworks would facilitate us to design and tune realistically adaptive synchronised polices for varying scenarios at specific time instances to fetch desired outcomes overcoming the non-desirable arbitrary impacts of the underlying influence networks.

\acknow{Please include your acknowledgments here, set in a single paragraph. Please do not include any acknowledgments in the Supporting Information, or anywhere else in the manuscript.}

\showacknow % Display the acknowledgements section

% Bibliography

\end{document}